# In Quest of an Extensible Multi-Level Harm Taxonomy for Adversarial AI: Heart of Security, Ethical Risk Scoring and Resilience Analytics*

Javed I. Khan [1] and Sharmila Rahman Prithula [1]

[1] Kent State University, Kent, OH 44242
javed@kent.edu | sprithul@kent.edu

**Abstract** *Harm* is invoked everywhere from cybersecurity, ethics, risk analysis, to adversarial AI—yet there exists no systematic or agreed-upon list of harms, and the concept itself is rarely defined with the precision required for serious analysis. Current discourse relies on vague, underspecified notions of harm, rendering nuanced, structured, and qualitative assessment effectively impossible. This paper challenges that gap directly. We introduce a structured and expandable taxonomy of harms, grounded in an ensemble of contemporary ethical theories, that makes harm explicit, enumerable, and analytically tractable. The proposed framework identifies 66+ distinct harm types, systematically organized into two overarching domains—human and non-human—and eleven major categories, each explicitly aligned with eleven dominant ethical theories. While extensible by design, the taxonomy's upper levels are intentionally stable. Beyond classification, we introduce a theory-aware taxonomy of victim entities and formalize normative harm attributes—including reversibility and duration—that materially alter ethical severity. Together, these contributions transform harm from a rhetorical placeholder into an operational object of analysis, enabling rigorous ethical reasoning and long-term safety evaluation of AI systems and other socio-technical domains where harm is a first-order concern.

**Keywords:** Artificial General Intelligence, Ethical analysis of AI, Cybersecurity, HARM66+

## 1 Introduction

Traditional cybersecurity harm models—such as the Confidentiality-Integrity-Availability (C-I-A) triad—are increasingly seen as insufficient for assessing risks posed by general-purpose AI [6,22]. With Artificial General Intelligence (AGI) approaching near-human levels of analysis and decision-making roles, these are now capable of producing harm that is diffused, cross-domain, and difficult to attribute. These harms extend well beyond isolated technical failures, affecting psychological well-being, institutional legitimacy, epistemic integrity, environmental systems, and even intergenerational social stability. The legacy frameworks were designed for bounded systems and fail to capture the expansive, cross-domain, and often deeply human impacts that advanced adversarial AI systems are now capable of producing. As such, there is a need for a broader, more universal harm model that reflects the scale, complexity, and scope of these emergent technologies [8].

Despite the centrality of harm in contemporary AI and technology ethics, there is still no widely accepted, comprehensive, and operationally usable typology of harms in the literature [15,19]. Existing classification schemes are often narrow, overly abstract, or lack the structural rigor required for systematic cross-domain ethical assessment, modular analysis, or quantitative scoring. Besides Adversarial AI, Harm classification is foundational and utterly central for too many domains [21].

This paper brings attention to this critical gap and introduces a structured and expandable taxonomy of harm. Grounded in an ensemble of contemporary ethical theories, we present a comprehensive set of 66+ distinct ethical harm types, systematically organized into two overarching domains and 11 categories. We call this set HARM66+. The categorization not only carefully includes the existing harms listed in contemporary literature but is also grounded in foundational support from major ethical theories, ensuring philosophical coherence and normative relevance. While the taxonomy is designed to be extensible, its upper levels—domains and categories—are intended to remain robust and stable, allowing broad applicability across contexts with minimal need for modification.

Harm analysis is inherently multidimensional; ethical evaluation depends not only on the type of harm, but also on the other associated characteristics – such as the entities involved—particularly the harmed entity, whose moral status is often central to ethical judgment. Similarly, ethical theories consider irreversibility and duration of harm. These factors are cornerstones for more nuanced and ethically grounded assessments of harm severity. We also discuss these with a view towards a quantitative framework for harm analysis.

Before we present the proposed classification, in section 2, we briefly discuss the related work in harm classification in existing literature. Section-3 provides the proposed classification of Ethical Harms. Section-4 discusses the type of entities who can harm and on whom harm can be inflicted. We also show how these entities can be weighed based on consideration of various ethical theories. In section 5 and 6 we then discuss the role of irreversibility and duration in harm analysis and integrating these dimensions (rubric) enables more nuanced, rigorous, and actionable harm assessment.

## 2 Related Works: Harm & Classification

In **bioethics and medical ethics** Harm is traditionally classified into five human (patient) only categories such as physical, psychological, social, legal [3] in the context of clinical trials and human subject research. In **environmental governance,** the Intergovernmental Panel on Climate Change (IPCC) focuses comprehensively on risk framework to assess climate-related harms [10,12]. It does not classify the harms.

**In the domain of Artificial intelligence** (AI), one of the most interesting works on harm is being pursued by the OECD. Their AI Principles and AI Now Institute frameworks identify a few digital harms such as ‘algorithmic bias’, ‘loss of autonomy’, ‘privacy violation’, and ‘systemic social injustice’ [2,20]. The Policy Observatory by OECD.AI is a living and growing list of 'AI incidents' defined as "*An event in which the use of AI systems resulted in outcomes that caused harm or raised substantial concerns about fairness, safety, transparency, or accountability*" [16]. The OECD categorizes AI-related events into four areas: 'rights and liberties', 'labor and automation', 'bias and inclusion', and 'safety and critical infrastructure'. [22] provided a typology of digital harms with three classes- ‘direct human’, ‘societal’, and a new class, ‘ontological harm’. He discovered that digital technology changes the nature and values of human interaction and society and named it as ontological harm.

**Table 1.** HARM66+ Taxonomy of Ethical Harms

| | Category | H-Code | Sub-subcategory | Category | H-Code | Sub-subcategory |
|---|---|---|---|---|---|---|
| Exo-Human Harms | Environmental and Ecological Harm | A.E1.01 | Harm to Animals | Economic and Corporate Harm | A.E4.01 | Fraud, counterfeit, embezellment |
| | | A.E1.02 | Ecosystem and Environmental Harm | | A.E4.02 | Business Ethics Violations, Unfair Trade Practice |
| | | A.E1.03 | Harm to Plants and | | A.E4.03 | Intellectual Property, Tradepractice |
| | | A.E1.04 | Disruption of Natural | | A.E4.04 | Financial System Harm |
| | | A.E1.05 | **Space and** | | A.E4.05 | Workforce and Employment Harm |
| | Digital and Technological Harm | A.E2.01 | **High Carbon & E-Waste** | | A.E4.06 | Legal and Regulatory Harm |
| | | A.E2.02 | AI and Automation Harm | | A.E4.07 | Oprational and Infrastrcture Harm |
| | | A.E2.03 | Cyber-Physical and | | A.E4.08 | Loss of Business Data |
| | | A.E2.04 | Emerging Technology | | A.E4.09 | Reputational and Trust Harm |
| | | A.E2.05 | Cognitive and Psychological Tech Harm | | A.E4.10 | Governance and Internal Harm |
| | | A.E2.06 | Neurotechnological Harm | Social, Cultural, & Political Harm | A.E5.01 | **Educational disruption learning** |
| | | A.E2.07 | Surveillance and Privacy | | A.E5.02 | Societal and Cultural Erosion |
| | | A.E2.08 | Exposure to cyberattack | | A.E5.03 | Truth and Knowledge Manipulation |
| | Physical Infrastructure Harm | A.E3.01 | Energy Infrastructure | | A.E5.04 | Heritage and Artistic Integrity Harm |
| | | A.E3.02 | Transportation | | A.E5.05 | Justice and Political System Harm |
| | | A.E3.03 | Communication System Failures | | A.E5.06 | Political Oppression and Authoritarian Control |
| | | A.E3.04 | Water and Sanitation | | A.E5.07 | **Loss of Sovereign Control** |
| | | A.E3.05 | Structural and Built Environment Failures | | A.E5.08 | Attack against Social Institutions |
| | | | | | A.E5.09 | Instigation of Social Violence |
| | | A.E3.06 | Public Safety System | | A.E5.1 | **Collapse of Epistemic Trust** |
| Endo-Human Harms | Physical or Medical Harm | H.H1.01 | Bodily Harm and Injury | Reputational or Identity Harm | H.H3.01 | Defamation and Public Shaming |
| | | H.H1.02 | Health and Disease Harm | | H.H3.02 | Loss of Trust and Social Standing |
| | | H.H1.03 | Displacement and Forced | | H.H3.03 | Identity Theft |
| | | H.H1.04 | Technology Induced Trauma and Torture | | H.H3.04 | Perosnality Theft |
| | Psychological & Cognitive Harm | H.H2.01 | Mental Health Harm | | H.H3.05 | Loss/Disclouse of medical privacy |
| | | H.H2.02 | Exploitation of Psychological Vulnerability | | H.H3.06 | Loss of social privacy |
| | | H.H2.03 | **Cognitive Manipulation and Behavioral Control** | | H.H3.07 | Loss of financial privacy |
| | Social or Relational Harm | H.H4.01 | Loss of Family and Community Bonds | | H.H3.08 | Tracking for political, social believe. |
| | | H.H4.02 | Breach of Trust and | | H.H3.09 | **Reputational Blackmail** |
| | | H.H4.03 | Social Isolation and | Financial or Occupational Harm | H.H6.01 | Loss of Money and Assets |
| | Political, Legal or Expressional Harm | H.H5.01 | Overuse of Power | | H.H6.02 | Job Loss and Economic Insecurity |
| | | H.H5.02 | Wrongful Legal Actions and Due Process Failures | | H.H6.03 | Fraud, Scams, or Financial Exploitation |
| | | H.H5.03 | Political Oppression, Legal Inequality and Injustice | | H.H6.04 | **Algorithmic Discrimination & Bias against Individual** |
| | | H.H5.04 | **Loss of Individual Control , Agency, and Choice** | | H.H6.05 | Increased Poverty and Economic Inequality |

**In cybersecurity**, harm and risk have traditionally been framed through the Confidentiality–Integrity–Availability (CIA) triad. While foundational, this model has long been criticized as overly simplistic. Parker’s expansion into the Parkerian Hexad introduced additional categories—authenticity, possession or control, and utility—to address limitations of the CIA [17]. Subsequent approaches, such as FAIR (Factor Analysis of Information Risk), further extended cybersecurity risk analysis by incorporating few secondary harms, including reputational damage and legal or regulatory losses [9].

Despite these extensions, cybersecurity frameworks remain largely asset-centric and lack a unified, ethically grounded taxonomy capable of capturing broader human, societal, and systemic harms. As AI systems become increasingly embedded in human decision-making and cyber-physical infrastructures, cyber risks can no longer be confined to technical failures alone but may propagate into nearly any class of classical harm affecting individuals, institutions, or society at large. As evident, most existing harm classification frameworks are i) have low in resolution (few broad abstract/harm types), ii) are abstract, iii) have very limited classification with ~1 level, and iv) often domain-specific. These are rarely structured for modular analysis, or scoring.

These longstanding limitations have been explicitly noted by leading scholars. Floridi et al. [8], for example, has argued that *“we must move beyond abstract principles and engage in systematic classification of harms—both known and emerging—to ensure meaningful ethical governance.”* Zuboff [25] has pointed out that the harms of digital systems are systemic, opaque, and under-classified, often escaping legal and ethical accountability because no detailed harm model exists to map how power, autonomy, and manipulation are affected. In ISO 31000 and NIST RMF documents [11], there is frequent mention that harm is “context-dependent” and must be decomposed into “specific, measurable impact domains.” Bostrom & Yudkowsky [4, 9], have drawn attention to AI’s catastrophic and existential risks that needed structured models of harm beyond physical death or system failure, including epistemic, societal, and intergenerational categories. The HARM66+ framework addresses this unmet need by providing a structured, extensible classification that many scholars have argued is essential.

# 3 Classification of Ethical Harm

## 3.1 Definition

E*thical harm* is defined as any adverse effect or damage inflicted upon individuals, groups, systems, or environments, either intentionally or not, from an action, event, or condition [3,25].

### 3.2 Classification

Table 1 provides the HARM66+ classification, a formal, multi-level taxonomy for analyzing harm in socio-technical systems. It organizes 66 distinct harm types into three tiers: (i) Major Harm Categories, which distinguish between Exo-Human Harms (impacting systems like environments and infrastructure) and Endo-Human Harms (affecting individuals); (ii) Intermediate Categories, which group harms into domains (e.g., environmental, technological, psychological); and (iii) Sub-subcategories, which specific harms like "Defamation," "AI Bias," or "Water System Failure." Each harm has a hierarchical identifier for granular reference and traceability.

The taxonomy is designed for mutual exclusiveness, orthogonality, normative completeness, hierarchy, stability, and parsimony. It maps all harm from 1500+ AI incidents as defined by the AI-AAIC Monitor [1]. Similarly, it maps 13 social risk categories identified from an expert review designed for **mutual exclusivity, orthogonality, completeness, hierarchical traceability, stability, and parsimony** of 1,781 documents [24]. HARM66+ was tested through a *normative non-reducibility test*: for each harm category, we evaluated whether removing that category would leave a class of ethically salient harms that could no longer be expressed without loss of moral meaning. If the removal of a category renders certain harms normatively inexpressible—i.e., irreducible to remaining categories without collapsing distinct ethical concerns—then the category is considered orthogonal. All categories in HARM66+ passed this test, indicating that each captures a distinct ethical dimension that cannot be subsumed by others without loss of normative resolution. Dotted coding allows for the extension of subclasses down to specific *incident instances*. The taxonomy was also stress-tested against futuristic harms portrayed in 16 major science fictions. While extensible, the taxonomy's upper levels (domains and categories) are designed for stability and broad applicability with minimal modification.

### 3.3 Support from Ethical Theories

Any harm taxonomy implicitly embeds value judgments about *what counts as harm* and *whose harm matters*. How do we ensure these are not ad hoc? Aligning HARM66+ with The 11 ethical theories ensure the taxonomy is **pluralistically defensible**, meaning harms remain valid even when ethical priorities differ. This is essential for cross-jurisdictional and cross-cultural legitimacy. We therefore validate the normative significance of HARM66+ from a set of ethical theories to ensure that selected harms are ethically grounded, pluralistically defensible, and resistant to claims of arbitrariness—transforming the taxonomy from a technical artifact into a normative instrument. Table 2(a) shows how major ethical theories[1] (MTEST11) view these harms, using codes for direct ("D"), indirect ("I"), conditional ("C"), or neutral ("N") consideration. All theories considered these harms unethical, with varied reasoning. For example, Psychological and Cognitive Harm is rated directly unethical ("D") by Utilitarianism (di-minishes well-being), Deontology (violates duties of respect), Virtue Ethics (reflects vice), and Rights-Based Ethics (violates mental integrity). Natural Law Theory offers only indirect ("I") support, as its primary focus is on bodily integrity. Using a Likert scale for quantitative assessment, Table 2(b) presents statistical analysis of this support. A strong consensus (Mean ≥ 2.0, p-value < 0.05) was found for nine harm categories, including Environmental, Economic, Physical (Human), and Psychological. Moderate consensus (Mean ≥ 1.5, p-value < 0.05) was found for Digital & Technological and Financial & Occupational harms. No categories showed weak or no consensus in this analysis.

---

[1] We consult 11 major ethical theories- a curated mix of normative and social ethics. The members have been widely applied in domains like healthcare, law, and governance. They include the classical triad—duty, outcome, and virtue—for individual ethics, plus social/political theories for societal (community) ethics. The set maintains internal orthogonality with purposeful over-lap for robust moral

**Table 2(a).** Views of Ethical Theories on Harms

| Harm ↓.Ethical Theory → | Harm Code | Utilitarianism | Deontology | Virtue Ethics | Ethics of Care | Rights-Based Ethics | Social Contract | Rawlsian Justice | Natural Law | Environmental Ethics | Pragmatism | Existentialist Ethics |
|---|---|---|---|---|---|---|---|---|---|---|---|---|
| Environmental and Ecological Harm | A.E1.00 | D | D | D | D | C | C | C | C | D | D | C |
| Digital and Technological Harm | A.E2.00 | I | D | C | I | C | C | C | C | I | D | C |
| Physical Infrastructure Harm | A.E3.00 | D | D | I | C | C | D | D | D | I | D | I |
| Economic and Corporate Harm | A.E4.00 | D | D | D | C | D | D | D | C | C | D | C |
| Social, Cultural, & Political Harm | A.E5.00 | D | D | D | D | D | D | D | C | I | D | D |
| Physical or Medical Harm | H.H1.00 | D | D | D | D | D | D | D | D | I | D | D |
| Psychological & Cognitive Harm | H.H2.00 | D | D | D | D | D | C | C | I | I | D | D |
| Reputational or Identity Harm | H.H3.00 | C | C | D | D | D | C | C | I | I | D | D |
| Social or Relational Harm | H.H4.00 | C | D | D | D | D | C | C | I | C | C | D |
| Political, Legal or Expressional Harm | H.H5.00 | D | D | D | C | D | D | D | C | C | C | D |
| Financial or Occupational Harm | H.H6.00 | D | C | D | C | D | C | D | C | C | C | D |

**Table 2(b).** View Analysis

| Mean | Std Dev | Z-Score | T-Statistic | P-Value | Consensus? |
|---|---|---|---|---|---|
| 2.09 | 1.04 | -0.43 | 6.64 | 5.78455E-05 | Strong |
| 1.64 | 0.81 | -1.68 | 6.71 | 5.31018E-05 | Moderate |
| 2.36 | 0.81 | 0.32 | 9.69 | 2.12054E-06 | Strong |
| 2.27 | 1.01 | 0.07 | 7.47 | 2.13478E-05 | Strong |
| 2.73 | 0.65 | 1.32 | 13.99 | 6.83006E-08 | Strong |
| 2.91 | 0.30 | 1.82 | 32.00 | 2.09052E-11 | Strong |
| 2.45 | 0.82 | 0.57 | 9.93 | 1.70241E-06 | Strong |
| 2.09 | 0.94 | -0.43 | 7.35 | 2.46206E-05 | Strong |
| 2.00 | 1.00 | -0.68 | 6.63 | 5.83015E-05 | Strong |
| 2.27 | 1.01 | 0.07 | 7.47 | 2.13478E-05 | Strong |
| 1.91 | 1.04 | -0.93 | 6.06 | 0.000121628 | Moderate |

# 4 Entities that Can Suffer Harm

## 4.1 Entity Classification

Existing technical work on victim classification also remains fragmented across domains such as cybersecurity, public health, AI ethics, and legal studies, with each field offering partial taxonomies tailored to its specific harm context. In cybersecurity, frameworks such as the MITRE ATT&CK Matrix [14] and ENISA threat models [7] typically categorize victims as end users, devices, or enterprise systems, emphasizing attack vectors and digital exposure rather than moral salience. In disaster risk and epidemiological ethics, classification schemes focus on human vulnerability, identifying populations by age, disability, income, or geography to prioritize resource

---

triangulation, making it ideal for handling the moral uncertainty of global technology. Real-world governance, regulators, courts, institutions, and cultures draw on different moral logics simultaneously (rights, utility, care, justice, environmental value). Aligning HARM66+ with the 11 theories ensures the taxonomy remain valid even when ethical priorities differ- essential for cross-jurisdictional and cross-cultural legitimacy.

**Table 3**. Entities that can be Subject to Harm

| | | Code | Entity Subclass | Description |
|---|---|---|---|---|
| Living Entitities | Individual Human | E1a | Living Individuals | Biological persons currently alive who can suffer the harms. |
| | | E1b | Deceased Individuals | Deceased persons whose identity, reputation, or digital legacy may be misused or harmed. |
| | Groups or Communities | E2a | Identity Groups | Collectives defined by race, ethnicity, or religion, or occupation. |
| | | E2b | Stakeholder Groups | People such as worker, employee, customers, patients, or buyers affected by product harms or misrepresentation. |
| | | E2c | Marginalized Groups | Vulnerable or Merginalized populations structurally excluded or harmed. |
| | | E2d | Digital Communities | Communities in common digital spaces. |
| | | E2x | Other Groups | Others groups or herds |
| | Institutions and Organizations | E3a | Public Institutions | Government bodies, agencies, or courts. |
| | | E3b | Formal Organizations | Corporations, companies, non-profits. |
| | | E3c | Jurisdictional Entities | Cities, counties, provinces, states, or nations- formal administrative or governing units. |
| | | E3x | Other Organizations | Other complex prganizations & ecosystems |
| | Non-Human Lifeforms | E4a | Animals | Sentient or domesticated animals affected by environment or systems. |
| | | E4b | Plants and Agriculture | Flora and crops damaged by pollution or technology. |
| | | E4x | Other lifeforms | Artificial/Syhthetic/ Alien/Extraterrestials Lifeforms |
| Inanimate & Abstract Entitities | Environmental and Natural Systems | E5a | Terrestrial Ecosystems | Forests, grasslands, mountains affected by deforestation, erosion, or climate events. |
| | | E5b | Aquatic Ecosystems | Oceans, rivers, lakes, wetlands harmed by pollution, acidification, or overfishing. |
| | | E5c | Atmospheric and Climate Systems | Air quality, climate, weather patterns altered by carbon emissions, aerosols. |
| | | E5d | Planetary and Extraterrestrial Systems | Outer space systems - planetary orbit, planetary atmospheres, celestial systems impacted by space debris and exploration. |
| | Built and Technological Entities | E6a | Physical Infrastructure | Tangible, material systems/assets — includes roads, pipelines, cables, water mains, buildings, satellites, towers, etc. |
| | | E6b | Built Systems | Engineered systems (digital or cyber-physical) — includes software, algorithms, control networks, data center, cloud. |
| | | E6c | Built Services | Electric utility, transportation, web services, emergency response, etc. Attack = disruption of availability, reliability, or trust in the service. |
| | Abstract and Social Constructs | E7a | Normative Justice Systems | Social structures for adjudicating fairness, accountability, and rights. |
| | | E7b | Epistemic Foundations | shared ethods for determining truth (e.g., scientific method, journalism, peer review). |
| | | E7c | Relational Trust and Legitimacy | The perceived legitimacy of actors and systems based on fair, honest, and consistent behavior. |
| | | E7d | Democratic and Civic Norms | Practices of voting, representation, civic engagement, and constitutional freedoms. |
| | | E7e | Faith, Belief, and Worldviews | Religious/spiritual beliefs, traditions, and ideological worldviews . |
| | | E7f | Linguistic and Cultural Expressions | Languages, oral traditions, artistic expressions, and cultural symbols that shape identity and communication. |
| | | | | (reserved) |

allocation and harm mitigation [5]. AI ethics scholarship has more recently expanded moral consideration beyond individual persons to include digital systems, institutional processes, and epistemic constructs such as truth, trust, or democracy [8]. Legal frameworks, including the UN Human Rights Conventions (UN, 1948) [23] and the European Convention on Human Rights [18], classify victims primarily by group identity or protected legal status, but typically exclude non-human entities, infrastructural systems, or abstract societal goods. Overall, these approaches lack either taxonomic granularity or explicit alignment with ethical theory. Understanding the victim is also fundamentally essential for evaluating harm within complex socio-technical systems. To address this gap, we also propose ENTITY7- a structured victim taxonomy spanning individuals, communities, institutions, non-human life, and abstract normative constructs to support systematic harm modeling. Table 3 presents the resulting classification of entities subject to ethical harm.

### 4.2 Victims & Ethical Theories

Victim classification plays a foundational role in ethical reasoning. Different ethical theories prioritize different classes of victims, leading to systematically divergent harm evaluations. Within the MTEST11 framework, utilitarianism (T1) assigns substantial moral weight to aggregate harms affecting groups (E2b) and ecosystems (E5a), whereas deontology (T2) and rights-based ethics (T5) emphasize harms to individuals (E1a) and protected or marginalized populations (E2a, E2c). Ethics of care (T4), virtue ethics (T3), environmental ethics (T9), and existentialist ethics (T11) variously prioritize vulnerability, relational dependence, non-human life (E4a), and individual agency (E1a). A robust and expressive victim taxonomy such as ENTITY7 is therefore essential for theory-aligned ethical reasoning. Table 4 summarizes how different ethical theories weight victim categories in harm evaluation.

**Table 4.** How Victim Class Impacts Ethical Evaluation in Major Ethical Theories

| Ethical Thoery | Victim Class | How Victim Classification Affects Moral Weight from the Theory |
|---|---|---|
| **Utilitarianism** | **E2b** (User Groups), **E5a** (Ecosystems) | Harms to large populations or ecosystems are morally weighty due to the cumulative suffering across many beings or systems. |
| **Deontological Ethics** | **E1a** (Living Individuals) | Harming individuals violates intrinsic moral duties—e.g., not lying, not killing—even if doing so may lead to beneficial outcomes. |
| **Virtue Ethics** | **E2c** (Marginalized Populations), **E7f** (Cultural Expression) | Ethical weight increases when harm reflects a lack of compassion or justice, especially toward the vulnerable or cultural heritage. |
| **Ethics of Care** | **E2c** (Marginalized Groups), **E1a** (Individuals) | Harms to those in dependent or care-based relationships are ethically severe due to the breach of relational responsibility. |
| **Rights-Based Ethics** | **E1a** (Individuals), **E2a** (Ethnic Groups) | Morally evaluates whether rights (privacy, expression, bodily integrity) are violated—prioritizing sentient or right-bearing beings. |
| **Social Contract Theory** | **E3a** (Public Institutions), **E7d** (Democratic Norms) | Harms to institutions or civic norms violate the agreed structure of society—undermining justice, law, or democracy for all. |
| **Rawlsian Justice** | **E2c** (Vulnerable Groups), **E3c** (Jurisdictions) | Justice prioritizes the most disadvantaged—harms to structurally oppressed or unrepresented populations are given moral urgency. |
| **Natural Law Theory** | **E4a** (Animals), **E7e** (Worldviews) | Harms that disrupt the natural order or purpose of life (e.g., procreation, family, nature) violate moral law, even if unintended. |
| **Environmental Ethics** | **E5a–E5d** (All Environmental Systems) | Assigns intrinsic moral value to non-human systems; harms to forests, oceans, or the climate are wrong even without direct human impact. |
| **Pragmatism** | **E6c** (Built Services), **E3b** (Organizations) | Ethics is judged by functional outcomes—if harm disrupts essential services or institutions, it's morally significant, regardless of theory. |
| **Existentialist Ethics** | **E1a** (Individuals), **E7c** (Relational Trust) | Suppression of autonomy, freedom, or personal authenticity are the gravest—especially when individuals are manipulated. |

### 4.3 Harmer's Role in Ethics

While the harmer's identity is relevant for assessing blame (especially in theories like virtue ethics or ethics of care), victim classification is more central to determining a harm's moral weight. It is foundational for ethical deliberation, clarifying who is morally considerable, how harm is counted under different theories, and how competing harms are weighed. Without it, ethical evaluation risks becoming arbitrary or biased toward visible or anthropocentric harm.

## 5 Irreversibility of Harm

### 5.1 Impact

The irreversibility of harm heightens its ethical severity because it removes the possibility of repair or compensation. In contrast, reversibility allows for moral recovery and accountability, mitigating a harm's ethical weight. Ethical and legal systems often consider reversibility to determine responsibility and the permissibility of risk. Table 5 described the irreversibility characteristics of the 11 harm categories.

### 5.2 Irreversibility and Ethical Theories

The significance of reversibility varies across major theories, each offering distinct perspectives on its moral weight. For Utilitarianism (T1), reversibility is highly important. Reversible harms are more acceptable if they yield greater net benefits, while irreversible harms carry more ethical weight due to their lasting negative impact. Deontological ethics (T2) prioritizes adherence to moral duties, making reversibility secondary. A morally wrong action remains impermissible even if the harm is undone, as the focus is on the intrinsic morality of the act itself, not its consequences. Virtue ethics (T3) views reversibility as contextually significant. A virtuous person assesses it, and causing irreversible harm may reflect a lack of virtues like prudence, whereas reversible harms allow for moral repair. The ethics of care (T4) finds irreversible harm particularly severe because they can irreparably damage relationships and trust. Reversible harms offer opportunities for care and repair, central to theory. Rights-based ethics (T5) considers reversibility a minor factor. The focus is on the inviolability of rights, so any violation is unethical regardless of whether the harm can be remedied. Social contract theory (T6) deems reversibility con-textually relevant. Rational agents would reject systems permitting irreversible harm, making reversible harm more tolerable under a fair social contract. Rawlsian justice (T7) assigns significant weight to irreversibility, especially for the least advantage. Irreversible harm violates fairness and cannot be offset, making reversibility crucial for long-term justice. Natural law theory (T8) disregards reversibility as a decisive factor. Actions are judged on their alignment with natural purpose, making them inherently right or wrong regardless of their consequences. Environmental ethics (T9) considers reversibility a core concern.

**Table 5**. Irreversibility & Durance Characteristics of Various Harm Types

| Harm Type | Irreversibility Rank | H-Code | Reversibility Description | Durance | Contextual Notes |
|---|---|---|---|---|---|
| Environmental and Ecological Harm | 1 | A.E1.00 | Often **irreversible** (e.g., extinction, ecosystem collapse). Timescales for recovery (if possible) can be centuries. | Long-Term | Ecosystem collapse, extinction, or climate shifts are typically irreversible or generational. |
| Physical Infrastructure Harm | 2 | A.E3.00 | **Partially reversible**, but often expensive and time-consuming (e.g., dam failure, bridge collapse). Loss of life can make it irreversible. | Short to Medium | Data breaches or software failures may be quickly patched; privacy violations may persist. |
| Physical or Medical Harm | 3 | H.H1.00 | Some permanent injuries or deaths are **irreversible**. Minor injuries may be healed, but not all. | Medium to Long | Some infrastructure can be rebuilt, but losses of life, heritage, or housing may be permanent. |
| Psychological & Cognitive Harm | 4 | H.H2.00 | **Partially reversible**, but often persistent (e.g., trauma, PTSD). Treatment varies in efficacy. | Short to Medium | Economic damage is often recoverable (e.g., via restructuring or bailout), but reputational damage may linger. |
| Reputational or Identity Harm | 5 | H.H3.00 | Difficult to undo fully, especially in digital age. Stigma may persist even after vindication. | Medium to Long | Cultural erosion, political disenfranchisement, or institutional distrust often span generations. |
| Social, Cultural, & Political Harm | 6 | A.E5.00 | **Intergenerational effects** can be long-lasting. Societal trust, culture, or institutions may take decades to rebuild. | Short to Long | Minor injuries heal, but chronic illness or death is irreversible. Durance varies significantly. |
| Digital and Technological Harm | 7 | A.E2.00 | Varies widely—loss of data or system compromise may be **reversible** via backups or patches, but impacts (e.g., surveillance) may not be. | Medium to Long | Trauma and mental illness may persist for years or lifetime, but recovery is possible for some. |
| Legal, Political, or Expressional Harm | 8 | H.H5.00 | Legal redress or public apology can partially reverse the harm, but **record or damage may persist**. | Medium to Long | Digital-age reputational damage may be semi-permanent despite apologies or retraction. |
| Social or Relational Harm | 9 | H.H4.00 | Personal relationships can sometimes be repaired, but full trust may not return. **Moderately reversible**. | Short to Medium | Relationships may recover, but deep betrayal or trust loss can endure. Context-dependent. |
| Economic and Corporate Harm | 10 | A.E4.00 | Financial and organizational damages are often **reversible** through compensation, restructuring, or legal remedy. | Short to Long | A mistaken arrest may be resolved quickly; political suppression or exile may last decades. |
| Financial or Occupational Harm | 11 | H.H6.00 | Most **reversible** by restitution, re-employment, insurance, or welfare mechanisms. Less lasting than other harms. | Short to Medium | Most financial harm is recoverable through compensation, jobs, or aid, but systemic exclusion can extend duration. |

Harms that permanently degrade ecosystems or cause extinction are particularly egregious, making prevention of irreversible damage a paramount ethical obligation. Pragmatism (T10) focuses on practical outcomes. Reversible harms are more tolerable because they can be corrected, while irreversible harms limit future options, making reversibility a practical safeguard. Existentialist ethics (T11) views reversibility as significant for moral responsibility. Irreversible harms permanently define an individual's actions and moral identity, while reversible harms allow for redemption.

In summary, theories like Utilitarianism, Pragmatism, and Rawlsian Justice treat reversibility as core or con-textual. In contrast, Deontology, Rights-Based Ethics, and Natural Law Theory largely minimize it, while Existentialist Ethics links it to moral responsibility.

Here is an example of how three theories will ask the question of reversibility differently for Environmental Harm from Deforestation. Utilitarianism would calculate long-term loss vs. gain. If forest regrowth is possible, the harm may be tolerable; if irreversible (biodiversity loss), it's more ethically troubling. Deontological approaches may oppose it if it violates duties (e.g., non-maleficence, stewardship, or duties to future persons), regardless of reversibility. While Environmental Ethics will condemn it especially when species extinction or ecosystem collapse occurs, which are irreversible. Pragmatism would ask- can restoration or rewilding undo the harm? If yes, some harm may be justified conditionally.

Table 6 shows how the eleven ethical theories consider the reversibility of harms using four levels: Core (C), where it is central to judgment; Contextual (X), where it depends on the case; Minor (M), where it has limited influence; and Irrelevant (I), where it is not factored or dismissed an ethically unimportant. Table 5 shows the potential duration of various harm types.

**Table 6.** Importance of Irreversibility of Harm to Various Ethical Theories

| Harm ↓.Ethical Theory → | Harm Code | Utilitarianism | Deontology | Virtue Ethics | Ethics of Care | Rights-Based Ethics | Social Contract | Rawlsian Justice | Natural Law | Environmental Ethics | Pragmatism | Existentialist Ethics |
|---|---|---|---|---|---|---|---|---|---|---|---|---|
| Environmental and Ecological Harm | A.E1.00 | C | M | X | C | M | C | C | C | C | C | X |
| Digital and Technological Harm | A.E2.00 | X | M | X | X | M | X | X | M | X | C | C |
| Physical Infrastructure Harm | A.E3.00 | X | M | X | X | M | C | C | M | X | C | X |
| Economic and Corporate Harm | A.E4.00 | C | M | X | X | X | C | C | M | I | C | X |
| Social, Cultural, & Political Harm | A.E5.00 | X | M | X | C | M | X | C | M | X | X | C |
| Physical or Medical Harm | H.H1.00 | C | M | X | C | M | X | C | M | I | C | X |
| Psychological & Cognitive Harm | H.H2.00 | C | M | X | C | M | X | X | M | I | C | C |
| Reputational or Identity Harm | H.H3.00 | X | M | X | C | M | X | X | M | I | X | C |
| Social or Relational Harm | H.H4.00 | X | M | X | C | M | X | X | M | I | X | X |
| Political, Legal or Expressional Harm | H.H5.00 | X | M | X | X | M | X | C | M | I | X | X |
| Financial or Occupational Harm | H.H6.00 | C | M | X | X | M | C | C | M | I | C | X |

## Durance of Harm

### 5.3 Impact

The duration of harm—how long a negative impact persists—is also another significant factor in its ethical evaluation. Longer-lasting harms carry greater ethical weight because their effects accumulate, increasing the moral urgency for prevention. The ethical score rises sharply when harm is both long-lasting and irreversible, as this undermines future moral agency, justice, and ecological integrity. The greater the duration, the higher the moral burden. Table-5 (column 5 & 6) and Fig-1 shows some probable *context* and corresponding sample durations of the eleven ethical harms, categorized as long-term (generational/permanent), medium-term (1-10 years, possibly reversible), and short-term (temporary with high recovery potential).

### 5.4 Durance and Ethical Theories

The ethical significance of harm duration varies across major ethical theories, reflecting diverse moral priorities. Several theories assign high ethical weight to harm duration. Utilitarianism (T1) is highly concerned with duration, as longer harms produce greater cumulative suffering. The ethics of care (T4) also gives high weight, as long-term harm threatens relational trust and care networks. Rawlsian justice (T7) strongly emphasizes duration, especially for persistent harms that limit opportunity for the least advantage. Environmental ethics (T9) considers duration a central concern, as harms like extinction and climate change span generations. A second group of theories moderately considers duration. Virtue ethics (T3) views it as a reflection of the agent's character and prudence. Rights-based ethics (T5) finds any rights violation wrong, but prolonged denial intensifies the moral breach. Social contract theory (T6) holds that persistent harms erode the legitimacy of a moral order. Pragmatism (T10) focuses on practical outcomes, making persistent and unresolved harms a greater ethical problem. Existentialist ethics (T11) sees enduring consequences as deepening the weight of an individual's moral responsibility and identity. Notably, two theories give low importance to duration. Deontological ethics (T2) is concerned with adherence to moral duties, making an action's wrongness independent of how long its consequences last. Similarly, natural law theory (T8) evaluates acts based on their alignment with natural purpose, not the duration of their effects. Figure 2 graphically illustrates the relative weights assigned to harms' duration by these major ethical theories.

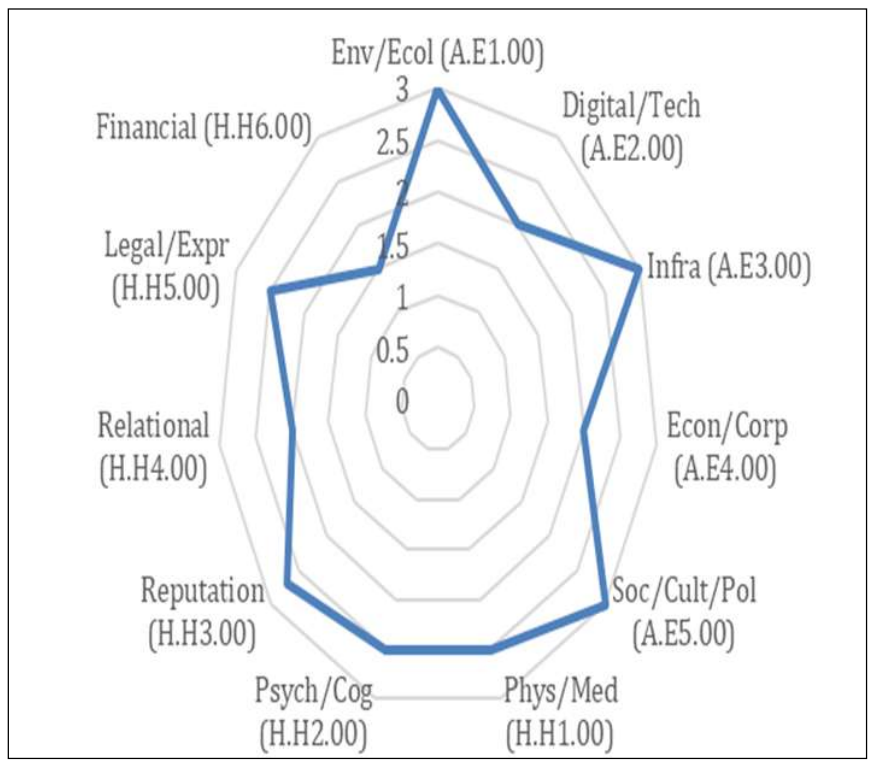

**Fig. 1.** Durance Scores of Various Harms (1=short, 2=medium, 3=long term)

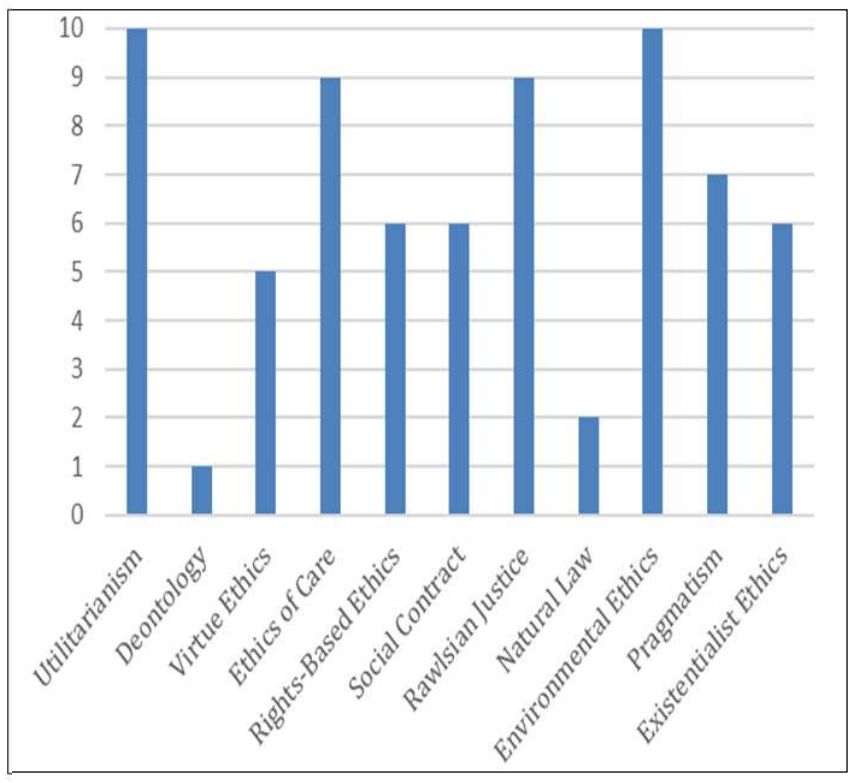

**Fig. 2.** Relative Importance of Durance to Various Ethical Theories

## 6 Use Cases for Taxonomy

A domain-level harm is a contextualized harm pattern observed within a specific socio-technical setting, while HARM66+ provides a domain-independent canonical harm vocabulary. The relationship between them is bidirectional. Forward mapping instantiates how a domain harm materializes as one or more HARM66+ harm classes. Reverse mapping uses HARM66+ as a diagnostic lens to identify harms that are present, latent, or unexamined within a domain. This two-way mapping supports both harm decomposition (domain → HARM66+) and harm completeness checking (HARM66+ → domain). The mapping is non-symmetric in strength: some HARM66+ harms are directly instantiated, others conditionally or weakly activated, while additional classes may remain structurally relevant even if not empirically observed.

Example: A domain harm such as loss of student agency due to automated decision systems directly instantiates Loss of Individual Control and Agency (H.H5.04), contributes to Cognitive Manipulation and Behavioral Control (H.H2.03), and conditionally activates Algorithmic Discrimination (H.H6.04). Reverse mapping surfaces additional concerns—such as Breach of Trust (H.H4.02) or Due Process Failures (H.H5.02)—that may not be explicitly named but warrant evaluation. This bidirectional framing enables harm analysis that is both context-sensitive and structurally complete, allowing comparison and mitigation across domains without collapsing domain-specific meaning.

Use Case A: When a university deploys AI for essay grading and placement, standard checks often focus on data quality, bias, and security. Mapping to HARM66+ reveals a host of additional risks such as loss of agency, psychological stress from opaque judgments, epistemic dependency, and erosion of credential trust—harms not visible through data-centric review alone.

Use Case B: For national oversight of AI in education, Education Commission applies HARM66+. It enables regulators to design much fine grained targeted, harm-specific measures—such as human-oversight requirements for agency loss or transparency obligations for epistemic harms—while providing a uniform vocabulary for standardized monitoring across institutions.

This taxonomy has also been applied to evaluate ethical risks in the data-harnessing processes of large language models (LLMs) [13].

# 7 Meta Harms

There is also notion of **meta-harms**—higher-order factors that shape the dynamics of harm without constituting harms themselves. Unlike first-order harms, which describe *what* is harmed, meta-harms describe *how* harms emerge, propagate, and intensify across time, scale, and interaction effects. We identify several – i) **Manifestation latency (**harms materialize only after prolonged or intergenerational delay). ii) **Feedback loops** and **harm compounding** (processes that initial harms reinforce themselves or interact with other harms, producing nonlinear escalation), iii) **Multiplicity** (capacity of a single system to generate multiple distinct harm types simultaneously across entities and domains), iv) **Scalability** characterizes harms that are individually minor but become severe when replicated at population or infrastructure scale. Finally, v) **invisibility and opacity** of harm. A taxonomy, such as HARM66+ can be the foundation for future analysis of such harm's dynamics.

# 8 Stability and Extensibility of Taxonomy

The taxonomy is designed to be both **stable and extensible**. Its upper levels—domains and major categories—are intentionally structured to remain robust over time, enabling broad applicability across socio-technical contexts with minimal modification. Extensibility is confined to lower-level subclasses, allowing the framework to accommodate emerging or context-specific harms without destabilizing its core structure. While the concept of harm is inevitably shaped by historical, cultural, and technological change, the taxonomy's foundational layers are intended to endure, even as terminology and subcategories evolve through continued scholarly refinement. To address concerns of universality, most importantly, HARM66+ has been grounded in **pluralistic ethical support** rather than a single normative doctrine. The taxonomy is affirmed across consequentialist, rights-based, care-oriented, justice-based, and environmental ethical perspectives, reducing the risk of cultural or temporal arbitrariness. This pluralistic grounding is essential for cross-jurisdictional and cross-cultural legitimacy, long-term relevance as priorities evolve over decade.

## 9 Conclusions

In this paper we call for addressing a critical gap in the literature by introducing a structured and expandable taxonomy of harm. We present HARM66+ classifying over 66 distinct harm types, systematically organized into two overarching domains and 11 categories. The categorization is grounded in foundational support from major contemporary ethical theories, ensuring philosophical coherence and normative relevance. We have also discussed its normative characteristics such as irreversibility and durance, and the classification of victim entities, as these will be weighted by major ethical theories. These classifications are expected to enable more nuanced and ethically grounded assessments of harm severity, supporting rigorous applications in both analytical modeling and policy decision-making. A specific, fine-grained, and standardized taxonomy provides three immediate benefits. Frameworks such as HARM66+ can remove ethical blind spots by surfacing latent harms—such as loss of agency, epistemic degradation, and institutional trust erosion—that compliance-oriented evaluations often miss. Second, their granularity enables harm-specific, proportionate mitigation, replacing coarse, ad hoc controls with safeguards aligned to harm severity and reversibility. Third, they establish a uniform cross-domain vocabulary, supporting consistent mitigation, comparable reporting, and interoperable data collection across institutions, sectors, and jurisdictions.

**Disclosure of Interests**: The authors declare that they have no competing interests that are relevant to the content of this article.